\documentclass[aps,prd,showpacs,twocolumn,superscriptaddress,nofootinbib,preprintnumbers]{revtex4-1} 

\makeatletter
\def\p@subsection{}
\makeatother

\usepackage{color,graphicx,amsmath,amssymb,lipsum}
\usepackage[colorlinks=true,citecolor=blue,linkcolor=blue,urlcolor=blue, backref=false,pdfborder={0 0 0}]{hyperref}
\usepackage{colortbl}
\definecolor{blue2}{cmyk}{1, 0.1, 0.1, 0}

\definecolor{pyBlue}{RGB}{31, 119, 180}
\definecolor{pyRed}{RGB}{214, 39, 40}
\definecolor{pyGreen}{RGB}{44, 160, 44}
\definecolor{pyBlue2}{RGB}{0, 111, 237}
\definecolor{pyRed2}{RGB}{224, 52, 36}

\usepackage{colortbl}
\definecolor{summersky}{cmyk}{0.71,0.33,0,0.5}
\definecolor{flamingo}{cmyk}{0,0.51,0.71,0.5}
\definecolor{rp}{cmyk}{0.2, 1, 0.6, 0}
\definecolor{pacificblue}{cmyk}{0.95,0.3,0, 0.5}
\definecolor{gray60}{cmyk}{0.4,0.4,0,0.8}

\newcommand{\red}[1]{\textcolor{pyRed}{#1}}
\newcommand{\blue}[1]{\textcolor{pyBlue}{#1}}

\usepackage[normalem]{ulem}
\usepackage[utf8]{inputenc} 
\usepackage{mathtools}
\usepackage{tabularx} 
\usepackage{tabu}
\usepackage{multirow}
\usepackage[table]{xcolor}
\usepackage{colortbl}
\usepackage{caption}
\usepackage{tikz}
\usepackage{bbm}
\usetikzlibrary{snakes}
\usetikzlibrary{decorations}
\usetikzlibrary{trees}
\usetikzlibrary{decorations.pathmorphing}
\usetikzlibrary{decorations.markings}
\usetikzlibrary{external}
\usetikzlibrary{intersections}
\usetikzlibrary{shapes,arrows}
\usetikzlibrary{arrows.meta}
\usetikzlibrary{calc}
\usetikzlibrary{shapes.misc}
\usetikzlibrary{decorations.text}
\usetikzlibrary{backgrounds}
\usetikzlibrary{fadings}
\usetikzlibrary{tikzmark,calc,arrows,shapes,decorations.pathreplacing}
\tikzset{
        cross/.style={cross out, draw=black, minimum size=2*(#1-\pgflinewidth), inner sep=0pt, outer sep=0pt},
	branchCut/.style={postaction={decorate},
		snake=zigzag,
		decoration = {snake=zigzag,segment length = 2mm, amplitude = 2mm}	
    }}

\usepackage{float}

\usepackage{bm}

\include{figures}

\usepackage{ulem}
\usepackage{diagbox}

\usepackage{mathrsfs}

\newcommand{\be}{\begin{equation}}
\newcommand{\ee}{\end{equation}}
\newcommand{\beqa}{\begin{eqnarray}}
\newcommand{\eeqa}{\end{eqnarray}}

\newcommand{\bseq}{\begin{subequations}}
\newcommand{\eseq}{\end{subequations}}

\renewcommand{\Im}{\mathop{\rm Im}\nolimits}

\newcommand{\eq}[1]{(\ref{#1})}

\definecolor{cornellRed}{HTML}{B31B1B}
\definecolor{cornellBlue}{HTML}{0068AC}
\definecolor{cornellGreen}{HTML}{6EB43F}
\definecolor{purple}{HTML}{66023C}

\def\gsim{\raise0.3ex\hbox{$\;>$\kern-0.75em\raise-1.1ex\hbox{$\sim\;$}}}
\def\lsim{\raise0.3ex\hbox{$\;<$\kern-0.75em\raise-1.1ex\hbox{$\sim\;$}}}

\def\beqn#1{\begin{equation}\label{#1}}
\def\eeqn{\end{equation}}

\def\beqa#1{\begin{eqnarray}\label{#1}}
\def\eeqa{\end{eqnarray}}

\def\Z2{$\mathcal{Z_2}$}

\newcommand {\ignore}[1]{}

\begin{document}

\title{5-Dimensional Gravitational Raman Scattering: Scalar Wave Perturbations in Schwarzschild-Tangherlini Spacetime}

\author{Samim Akhtar}
\email{samimakhtar@imsc.res.in}
\affiliation{The Institute of Mathematical Sciences,
 	IV Cross Road, C.I.T. Campus, Taramani, Chennai 600 113, India}
\affiliation{Homi Bhabha National Institute, 
	Training School Complex, Anushakti Nagar, Mumbai 400 094, India}
\author{Yilber Fabian Bautista}
\email{yilber-fabian.bautista-chivata@ipht.fr}
\affiliation{Institut de Physique Théorique, CEA, Université Paris–Saclay,
F–91191 Gif-sur-Yvette cedex, France}
\author{Cristoforo Iossa}
\email{Cristoforo.Iossa@unige.ch}
\affiliation{Section de Math\'ematiques, Universit\'e de Gen\`eve, 1211 Gen\`eve 4, Switzerland and\\ Theoretical Physics Department, CERN, 1211 Geneva 23, Switzerland}
\author{Zihan Zhou}
\email{zihanz@princeton.edu}
\affiliation{Department of Physics, Princeton University, Princeton, NJ 08540, USA}

\begin{abstract} 
In this Letter, we study scalar wave perturbations of arbitrary frequency to the  5D Schwarzschild-Tangherlini black hole (STBH) within general relativity. For the first time, we derive a closed formula for the 5D partial wave gravitational Raman scattering amplitude applicable to a broad class of boundary conditions, expressed in terms of the Nekrasov-Shatashvili (NS) function for the reduced confluent Heun problem. Furthermore, up to  $\mathcal{O}(G^2)$ we compute the dynamical $\ell=0$, and the static $\ell=1$, scalar tidal Love numbers of the STBH  by matching an effective field theory description for a scalar wave scattering off the black hole, to our novel ultraviolet-NS solutions. The matched Love numbers do not vanish and  present renormalization group running behavior.

\end{abstract}

\maketitle

\textit{Introduction}-- The direct detection of gravitational waves has opened a new window into the study of compact objects in the strong gravity regime \cite{LIGOScientific:2014pky,VIRGO:2014yos,LIGOScientific:2016aoc,LIGOScientific:2017vwq,LIGOScientific:2018mvr,LIGOScientific:2020ibl,LIGOScientific:2021usb,LIGOScientific:2021qlt,KAGRA:2020agh,KAGRA:2021vkt}. Due to the complexity in finding exact solutions to   Einstein's equations,  perturbative methods are used to model gravitational wave signals and understand how compact objects, such as black holes (BHs) and neutron stars, respond to external perturbations. In  effective field theory (EFT) setups \cite{Goldberger:2004jt,Goldberger:2022rqf,Porto:2016pyg,Kalin:2020mvi,Cheung:2018wkq,Kosower:2018adc,Bern:2020uwk,Mogull:2020sak}, compact objects are modeled as point particles with additional multipole moments that characterize their tidal deformability. The coefficients of these moments, known as tidal Love numbers, play a crucial role in distinguishing between different types of compact objects. Remarkably, in 4D general relativity (GR), the static  Love numbers of BHs vanish identically \cite{Binnington:2009bb,Damour:2009vw,Cardoso:2017cfl,Combaluzier-Szteinsznaider:2024sgb,Goldberger:2022ebt,Porto:2016zng,Charalambous:2021kcz,Hui:2021vcv,Charalambous:2022rre,Ivanov:2022qqt}. 

Recently, the tidal response of compact objects in GR has been studied by matching the 4D gravitational Raman scattering amplitudes in black hole perturbation theory (BHPT) \cite{Ivanov:2022qqt,Bautista:2023sdf,Saketh:2023bul,Bautista:2021wfy,Bautista:2022wjf}, the ultraviolet (UV) description, to the gauge-independent EFT computations \cite{Ivanov:2022qqt,Ivanov:2022hlo,Saketh:2023bul,Saketh:2024juq, Ivanov:2024sds,Caron-Huot:2025tlq}.  This has unambiguously proven that scalar static Love numbers of the  Schwarzschild BH indeed vanish,  and that the dynamical (time-dependent) Love numbers present renormalization group (RG) running features. Central to these developments is the modern understanding of  BHPT in terms of a special class of functions, the  Nekrasov-Shatashvili (NS) functions \cite{Nekrasov:2009rc,Aminov:2020yma}. This class of functions appeared for the first time in the context of four dimensional supersymmetric gauge theories \cite{Nekrasov:2002qd} and, via the AGT correspondence \cite{Alday:2009aq}, two dimensional conformal field theories.
This recent approach, discussed for the first time in \cite{Aminov:2020yma}, has uncovered important features of the gravitational Raman amplitude, such as factorization in the partial wave basis, BH spin-analytic properties \cite{Bautista:2023sdf},  quasi-normal mode (QNM) quantization conditions, and resummation of tail-contributions in binary gravitational waveforms \cite{Ivanov:2025ozg,Fucito:2024wlg}.

Despite substantial advances~\cite{Kol:2011vg,Hui:2020xxx,Glazer:2024eyi,Rodriguez:2023xjd,Charalambous:2025ekl,Gray:2024qys,Charalambous:2024tdj}, the tidal responses of higher-dimensional BHs remain largely unexplored, primarily due to the absence of systematic analytic solutions to the equations of higher-dimensional BHPT in generic frequency setups. In this Letter, we present the first systematic study of the gravitational Raman scattering amplitude resulting from massless scalar perturbations to the 5D Schwarzschild-Tangherlini black hole (STBH), using a  modern approach to BHPT.

Our key contributions in this Letter are:
\begin{itemize}
    \item Demonstration that the radial perturbation equation corresponds to the reduced confluent Heun equation (RCHE), whose connection coefficients can be elegantly put in terms of a NS-function.
    \item Derivation of an exact formula for the 5D partial wave gravitational Raman scattering amplitude for a generic class of boundary conditions (BCs), which can be evaluated to arbitrary order in powers of $(r_{s,5}\omega)$,  the post-Minkowskian  (PM) expansion.
    \item The first complete matching of the 5D dynamical $\ell=0$, and the static $\ell=1$  elastic tidal Love numbers of the STBH at $\mathcal{O}(G^2)$, and the leading $\ell=0$ dissipative Love number at $\mathcal{O}(G^{\frac{3}{2}})$. 
\end{itemize}

In the remaining of this work, we detail the derivation of the listed contributions, discussing the implications of the exact new solutions 
including the eikonal limit,  exploring the full matching with worldline EFT computations, elucidating the RG running of tidal Love numbers (even in the static case), and highlighting the associated tidal anomalous dimensions. In the supplementary material, we include additional details on the derivation of several results included in the main text.

\textit{Scalar Perturbations of the 5D STBH}--
Consider the 5D STBH  metric  \cite{Tangherlini:1963bw},
\begin{align}\label{eq:metric}
    ds^2 {=} {-}f(r) dt^2 {+} \frac{dr^2}{f(r)}  {+} r^2 d\Omega_{3}^2 \,,\,\,
    f(r) {=} 1{-}\left(\frac{r_{s,5}}{r}\right)^{2}\,,
\end{align}
where the Schwarzschild radius  $r_{s,5}=  \sqrt{\frac{8GM}{3 \pi }} $, with $M$ the BH mass and $G$ the 5D  Newton's constant. Scalar perturbations of this metric are studied by considering the wave equation for a massless scalar field $\phi(x)$, propagating in such a BH background. Separating the angular from the radial motion through $\phi(x)=\int_\omega e^{i\omega t}\sum_{\ell,m} Y_{\ell,m}(\theta) \psi(r)$,  with  $Y_{\ell,m}(\theta)$ the hyperspherical harmonics, (see e.g. Ref.~\cite{Hui:2020xxx,Glazer:2024eyi} ) the radial dynamics of the scalar field is  governed by the differential equation
\begin{equation}\label{eq:diffgend}
    f(r)\psi''(r) +f'(r)\psi'(r)+\Big(\frac{\omega^2-V(r)}{f(r)}\Big)\psi(r)=0\,,
\end{equation}
where $\omega$ is the wave frequency, and the  scalar potential  
 \begin{equation}
     V(r){=} f(r)\Big[  \frac{3 f'(r)}{2 r}{+}\frac{\ell (\ell{+}2) }{r^2}{+}\frac{ 3 f(r)}{4 r^2}\Big]\,, \,\,  \ell {=} 0,1,2,\cdots\,.
\end{equation}

The differential equation  \eqref{eq:diffgend} possess two regular singular points,  $r=0$ and $r=r_{s,5}$, and one irregular singular point, $r=\infty$, of Poincaré rank  $h=\frac{1}{2}$; these in turn correspond to the singularities of the reduced confluent Heun equation (RCHE) \cite{Ronveaux1995-am}. 
Introducing the change of variables
\begin{align}\label{eq:yvar}
    z {=} \left(\frac{r}{r_{s,5}}\right)^{2},\quad f{=} 1{-}z^{-1},\quad\psi(z) = w(z) z^{\frac{1}{4}}(z-1)^{-\frac{1}{2}}\,,
\end{align}
Eq. \eqref{eq:diffgend} indeed reduces to  the RCHE in the normal form 
\begin{equation}\label{eq:rcheun}
     \Big(  \partial_z^2 +\frac{u-\frac{1}{2}+a_0^2+a_1^2}{z(z-1)}+\frac{\frac{1}{4}-a_1^2}{(z-1)^2}+\frac{\frac{1}{4}-a_0^2}{z^2}-\frac{L^2}{4 z}\Big) w(z)=0\,,
\end{equation}
where the regular  singularities are mapped to the points  $z=0$ and $z=1$ respectively, and the irregular one to $z=\infty$.  The dictionary of parameters is 
\begin{equation}\label{eq:dictionary} 
\begin{split}
    a_0^2=0\,,\,\, a_1^2{=}\frac{L^2}{4}={-}\frac{1}{4}( r_{s,5} \omega)^2\,,\,\,
    u{=}\frac{1}{2}(r_{s,5} \omega)^2{-}\frac{1}{4} \ell (\ell+2)\,.
    \end{split}
\end{equation}

Near the BH horizon ($z\to1$),  Eq.~\eqref{eq:rcheun} possess two independent solutions,  $w(z) \simeq (1-z)^{\frac{1}{2}\pm a_1}$,  corresponding to   purely incoming ($+1$) or purely outgoing  ($-1$) wave-modes.  
     Scalar perturbations in a BH background should satisfy incoming boundary conditions (ibcs). It is however interesting to probe the effects of different boundary conditions  on the Raman amplitude \cite{Frolov:1998wf,Goldberger:2019sya}. 
  With this in mind we consider the more generic semi-reflective boundary conditions (s-rbcs.) formed by the superposition,  $w(z) \simeq A (1-z)^{\frac{1}{2}+ a_1} + B (1-z)^{\frac{1}{2}- a_1}$, with $A,B \in \mathbb{R}$.   A given choice of BCs at the horizon is then propagated to  $r\to\infty$ through a connection formula, a coordinate independent, gauge invariant object. For real wave frequencies, this results in a superposition of an incoming and a reflected wave at future null infinity\footnote{The requirement for real frequencies can be lifted in such a way that the incoming wave at $r=\infty$ in Eq.~\eq{eq:radial_asym} has vanishing amplitude. Such a requirement provides a quantization condition for the eigenfrequencies and eigenfunctions of the radial problem, known as the quasinormal mode quantization condition \cite{Berti:2009kk}.
}. This  dictates to take for every $\ell$-mode\footnote{The factor of  $(-1)^{\ell+3/2}$ comes from  $\psi_0(r)$, the  free ($G=0$) solution to the wave equation \eqref{eq:diffgend}, which is a superposition of Riccati-Hankel functions \cite{Caron-Huot:2025tlq}.  As $r\to\infty$,  $\psi_{0}(r) \to e^{\pm i(\omega r +\frac{\pi}{2}(\ell+\frac{3}{2}))} $.}

\begin{align} \label{eq:radial_asym}
    \psi_\ell(r)& =   A e^{-  i\omega r^\star}+B e^{ i\omega r^\star}\,, & r^{\star} \rightarrow -\infty \,, \cr
    \psi_\ell(r)& =  B^{\rm inc}_{5,\ell} \frac{e^{-i\omega r^\star}}{r^{\frac{3}{2}}} {+}  \frac{ B^{\rm ref}_{5,\ell}}{ (-1)^{-\ell{-}\frac{3}{2}}} \frac{e^{i\omega r^\star}}{r^{ \frac{3}{2}}} \,, &r^{\star} \rightarrow \infty \,,
\end{align}
where the tortoise coordinate $r^\star$ is obtained by solving $      \frac{d r^\star}{d r} = \frac{1}{f(r)}\,. $  The constants in \eqref{eq:radial_asym} are related to each other via the so called connection matrix, computed for the RCHE for the first time in \cite{Bonelli:2022ten}. The five-dimensional partial wave gravitational Raman amplitude, which is composed of an elastic phase-shift $\delta_{5,\ell}$ and a dissipation  parameter $\eta_{5,\ell}$, is defined from the ratio of the asymptotic coefficients:
\begin{equation}\label{eq:deffdelta}
    \eta_{5,\ell}e^{2i\delta_{5,\ell}} \equiv (-1)^{\ell+\frac{3}{2}}\frac{B^{\rm ref}_{5,\ell}}{B^{\rm inc}_{5,\ell}}\,.
\end{equation}

This ratio can be obtained in a closed form --- to all orders in the PM expansion --- using the modern formulation of  BHPT and the NS functions \cite{Dodelson:2022yvn,Aminov:2023jve,Bianchi:2021xpr,Bianchi:2022qph,Bianchi:2023sfs,Giusto:2023awo,Aminov:2020yma,Bonelli:2021uvf,Consoli:2022eey,Fucito:2023afe,DiRusso:2024hmd,Bianchi:2024vmi,Bianchi:2022qph,Cipriani:2024ygw}. Using the BCs given by Eq.~\eqref{eq:radial_asym}, we find the closed formula (see the supplementary material for a detailed derivation)

\begin{widetext}
\begin{equation}\label{eq:Bratio}
\boxed{
    \eta_{5,\ell} e^{2i\delta_{5,\ell}} ={\red{e^{i\pi (2a +\ell+1)}}} \times {\blue{\frac{1 + e^{-2i\pi a} \mathcal{K}_5}{1 + e^{2i\pi a}\mathcal{K}_5}
\left[ \frac{1 +  \mathcal{R}\frac{B}{A}\left(\frac{1 + e^{-2i\pi a} \mathcal{K}_{5}|_{a_1 \rightarrow-a_1}}{1 + e^{-2i\pi a} \mathcal{K}_{5}} \right)}{1 +  \mathcal{R}\frac{B}{A}\left(\frac{1 + e^{2i\pi a} \mathcal{K}_{5}|_{a_1 \rightarrow-a_1}}{1 + e^{2i\pi a} \mathcal{K}_{5}} \right) }  \right]}} \,.}
\end{equation}
Here, $\mathcal{R}$ is a combination of Gamma functions given in Eq.~\eqref{eq:Rgammas}, $\mathcal{K}_5$ is the five-dimensional BH tidal response function 

\begin{align}\label{eq:response5d}
    \mathcal{K}_5 &= \vert L \vert^{-4a} 2^{4a} \frac{\Gamma(2a) \Gamma(1+2a)  \Gamma(\frac{1}{2}+ a_1 + a_0 - a) \Gamma(\frac{1}{2}+ a_1 - a_0 - a)}{\Gamma(-2a) \Gamma(1-2a)  \Gamma(\frac{1}{2}+ a_1 + a_0 + a) \Gamma(\frac{1}{2}+ a_1 - a_0 + a)} e^{\partial_a F} \,,
\end{align}
\end{widetext}
and we have introduced  $F(a_0,a_1,L)$, the NS function  for a 4D, $\mathcal{N}=2$ supersymmetric gauge theory  with two mass hypermultiplets \cite{Nekrasov:2002qd,Nekrasov:2009rc}. $F$ admits a convergent expansion in terms of the instanton parameter $L$ defined in Eq.~\eqref{eq:dictionary}.  In the supplementary material we have included the result of the explicit evaluation of $F$ up to order $L^6$, although generic $L$-results can be algorithmically obtained. 
The parameter $a$ introduced in Eq.~\eqref{eq:Bratio} is mathematically known as the Floquet exponent of the radial perturbation equation. It can be obtained from the recursive solution to the  Matone relation \cite{Matone:1995rx,Flume:2004rp} 
\begin{equation}\label{eq:matone}
    u= \frac{1}{4}-a^2+\frac{L}{2}\partial_L F ~,
\end{equation}
which, in the PM expansion evaluates to 
\begin{widetext}
\begin{equation}\label{eq:aPM}
    a  =-\frac{\ell{+}1}{2}{+}\frac{3 (r_{s,5}\omega) ^2}{8 (\ell{+}1)} + \frac{5 \ell  (\ell +2) (7 \ell  (\ell +2)-17)-48}{128 (\ell -1) \ell  (\ell +1)^3 (\ell +2) (\ell +3)} (r_{s,5} \omega)^4 {+} \mathcal{O}(r_{s,5}\omega) ^6  ~.
\end{equation}
\end{widetext}
Physically, $a$ is an analog of the ``renormalized" angular momentum introduced in the solution to the Teukolsky equation \cite{Bautista:2023sdf}, which in 4D has been proved to coincide with the  anomalous dimension of the   multipole operators in GR \cite{Ivanov:2025ozg}. Note that $F$, $u$, and therefore $a$,  are invariant under independent sign flip of the parameters $\{a_0,a_1,L\}.$ The explicit evaluation of  $a$  up to $\mathcal{O}(L^6)$  is given in Eq.~\eqref{eq:aL6}, and  using the dictionary in Eq.~\eqref{eq:dictionary}  we obtain the PM-expanded version for $a$ shown in Eq.~\eqref{eq:aPM}.

We have therefore provided all the elements needed for the explicit evaluation of $\delta_{5,\ell}$ and  $\eta_{5,\ell}$, at any order in the $(r_{s,5}\omega)$-expansion,  for a generic class of boundary conditions,  thus greatly extending the traditional GR  computations obtained previously only in the static limit \cite{Kol:2011vg,Hui:2020xxx}.

Remarkably, the connection formula in Eq.~\eqref{eq:Bratio} shows a clear factorization of contributions from the near zone and the far zone physics (blue-red respectively). This is analogous to the 4D near-far factorization proposed in \cite{Ivanov:2022qqt,Bautista:2023sdf}, but shows great simplicity, especially in the far zone piece (compared with Eq.~(2.8) in Ref.~\cite{Bautista:2023sdf}). This is because  in 5D there is no Newtonian phase associated to  IR divergences  due to  long-range $1/r$ exchanges. 

In the perturbative PM-expansion,  the $a$ parameter in Eq.~\eqref{eq:matone} and the tidal response function in Eq.~\eqref{eq:response5d} develop  integer $\ell$-divergences, as evident already at  $\mathcal{O}(r_{s,5} \omega)^4$ in Eq.~\eqref{eq:aPM}.
These divergences are spurious and cancel once near and far zone contributions are added together, thus leaving finite the physical gravitational Raman scattering phase shift\footnote{In the analogous  $4$D case, both integer and half-integer $\ell$-divergences are present both in near zone and far zone, but precisely cancel in the total phase-shift (see Eq.~(2.8) in Ref.~\cite{Bautista:2023sdf}).}.

More formally, in an EFT setup, the appearance of spurious $\ell$-poles can be understood from the complex angular momentum theory of the perturbative  S-matrix \cite{Gribov:2003nw}. Such poles will, in general, appear when using the Froissart-Gribov formula to perform analytic continuation of the S-matrix in complex angular momentum space \cite{Caron-Huot:2017vep,Caron-Huot:2020nem,Ivanov:2024sds}. This indicates the need to introduce new local counterterms around those poles to remove such divergence. As we show below, in the EFT for scalar wave scattering off the STBH the $\ell=0$ and $\ell=1$ poles at $\mathcal{O}(r_{s,5}\omega)^4$ above signal the need for introducing a static tidal Love number, $c_{\phi,1}$, for the $\ell=1$ sector, and a dynamical tidal Love number, $c_{\omega^2\phi,0}$, for the $\ell=0$ sector. The residue of the pole naturally provides the $\beta$-function for the RG running of these tidal Love numbers.

It is illustrative to understand the leading-order   PM contribution to the Raman amplitude from the near-zone, for a given $\ell$-mode. 
The   tidal function  evaluates  to
\begin{align}\label{eq:leatink5}
    \mathcal{K}_5^{\text{LO}} {=}\left(r_{s,5}\omega\right)^{2(\ell{+}1)}  \frac{\Gamma \left({-}\frac{\ell+1}{2} \right) \Gamma \left(\frac{1{-}\ell}{2}\right)}{2^{6 (\ell{+}1)} \Gamma \left(\frac{\ell{+}1}{2}\right) \Gamma \left(\frac{\ell{+}3}{2}\right)}{+}\mathcal{O}(r_{s,5}\omega)^{2\ell{+}4}\,,
\end{align}
such that \footnote{Notice that although $\sin(\ell\pi)$ vanishes for $\ell$-odd, $\mathcal{K}_5$ diverges due to the odd-integer  $\ell$-poles of the gamma functions in the numerator of \eqref{eq:leatink5}. }
\begin{align}
  &  \frac{1{ +} e^{-2i\pi a} \mathcal{K}_5}{1 {+} e^{2i\pi a}\mathcal{K}_5}\Big[\cdots\Big]{=}1-2 i \sin (\pi  \ell)\mathcal{K}_5^{\text{LO}} +\mathcal{O}(r_{s,5}\omega)^{2\ell{+}4}\,,\nonumber\\
    &{=}
\begin{cases}
1 +\mathcal{O}(r_{s,5}\omega)^{2\ell{+}4} & \text{for}\, \ell\in \text{even}-\mathbb{Z}^{+}\,,\\
1{+}\frac{i \pi  8^{-2 n-1} (r_{s,5}\omega)^{2 n{+}2}}{(n-l) \Gamma \left(\frac{n+1}{2}\right)^2 \Gamma \left(\frac{n+3}{2}\right)^2}{+} \mathcal{O}(l{-}n)^0 & \text{for}\, n\in \text{odd}-\mathbb{Z}^{+}\,.
\end{cases}
\label{eq:tildal5D}
\end{align}
This suggests that all of the static Love numbers of even-$\ell$-harmonics  should vanish in 5D, as has also been observed in other static BHPT  computations  \cite{Kol:2011vg,Hui:2020xxx}. The odd $\ell$-numbers on the other hand would not vanish and present singularities that mix near and far zone contributions.  It is also interesting to note that Eq.~\eqref{eq:tildal5D} is independent of the type of boundary conditions imposed at the BH horizon, therefore suggesting a kind of universality of the static $\ell$-Love number at leading PM-order.

Before we move to study the details on an EFT Love matching, we provide an exact formula for the $a$-parameter in the eikonal limit: $r_{s,5} \omega \gg 1, \ell \gg1$, 
with $x\equiv (r_{s,5} \omega)/\ell = r_{s,5}/b$ fixed, and $b$  the impact parameter:
\begin{equation}
\begin{aligned}
\label{eq:eikonal}
    \frac{2a {+} \ell {+} 1}{\ell} & {=} \frac{3}{2} x^2 \, _3F_2\left(\frac{1}{2},\frac{5}{4},\frac{7}{4};\frac{3}{2},2;4 x^2\right) {-}\frac{2K\left(\frac{4 x}{2 x{+}1}\right)}{\pi  \sqrt{2 x{+}1}}{+}1~.
\end{aligned}
\end{equation}
Here $K$ is the elliptic integral of the first kind. This expression has a branch cut at $x=\pm 1/2$, which coincides with the position of the 5D BH shadow $b=2r_{s,5}$.
Moreover, the far zone scattering phase-shift can be greatly simplified in the high energy limit when $|r_{s,5} \omega| \gg 1$, since 
\begin{equation}
    a \simeq - \frac{i}{2} (r_{s,5} \omega) ~,
\end{equation}
as can be seen by taking the $x\rightarrow \infty$ limit of Eq.~\eqref{eq:eikonal}. More elegantly, this value arises from a great simplification of the NS function in this limit, which can be understood by exploiting the properties of the classical conformal blocks (see the supplementary material).

\textit{EFT  for the  STBH and  UV matching}--
We now study an EFT description for the scattering of a scalar wave off the STBH. We compute the effective retarded two point amplitude, i.e.~${\rm LSZ}\langle \phi_+\phi_-\rangle$, in the partial wave basis up to order $G^2$,
 where $\phi_\pm$ are massless scalar fields in the   Keldysh
basis \cite{Keldysh:1964ud,PhysRevA.8.423} representing the classical field and the fluctuations respectively (see Ref.~\cite{Caron-Huot:2025tlq} for details). Such an EFT amplitude is then 
matched to the UV gravitational Raman amplitude \eqref{eq:Bratio}, which allows us to unambiguously fix the full BH Love numbers discussed above.
 
The long-distance dynamics for the scattering problem is governed by the effective action 
\begin{equation}\label{eq:EFTaction}
    S_{\text{minimal}} = -\frac{1}{2} \int d^D x \sqrt{-\bar g} \bar g^{\mu\nu}\partial_\mu \phi \partial_\nu \phi\,,
\end{equation}
where  $\bar g$ is the STBH background metric, which we specialize to $D=5$.
Up to order $G^2$, we supplement the minimal action \eqref{eq:EFTaction} with the  non-minimal static $\ell=1$, and dynamical  $\ell=0$, worldline operators 
\begin{equation}\label{eq:Stidal}
    S_{\text{tidal}} = \frac{c_{\phi ,1}}{2}\int d\tau (\partial_i \phi)^2 + \frac{c_{\omega^2\phi ,0}}{2}\int d\tau \dot{\phi}^2\,,
\end{equation} 
where $c_{\omega^2\phi ,0} $ and $c_{\phi ,1}$ are bare coefficients needed  to absorb UV divergences of the 1-loop contributions  from the minimal action \eqref{eq:EFTaction}, 
and the dissipative  Schwinger-Keldysh   $\ell=0$ action \cite{Caron-Huot:2025tlq},
\begin{equation}\label{eq:dissipact}
    S_{\text{dissip.}} = c_{\omega \phi,0}  \int d\tau \phi_+\partial_\tau \phi_-\,,
\end{equation}
 needed to match the leading dissipative contribution in the Raman amplitude  \eqref{eq:Bratio}. The spatial and time derivatives are defined through the timelike vector $u^\mu=(1,0,0,0,0)$, via $\partial_i =(\bar g^{\mu\nu} + u^\mu u^\nu)\partial_\mu \ $, and $\dot{\phi}  = \partial_\tau \phi = u^\mu \partial_\mu \phi$ respectively. 

The computation of the retarded 1-loop amplitude is carried out using the background field method \cite{Ivanov:2024sds}, with the details provided in the supplementary material. 
Using the exponential representation of the scattering operator $S^{\text{EFT}} = e^{i\Delta^{\text{EFT}}}$, up to order $G^2$ the scattering phase  $i\Delta^{\text{EFT}}$ receives the contribution from the diagrams 
\begin{widetext}
\begin{equation}
\label{eq:EFT_Amp}
    i\Delta^{\text{EFT}} =  \begin{gathered}
    \begin{tikzpicture}[line width=1,photon/.style={decorate, decoration={snake, amplitude=1pt, segment length=6pt}}]
    \draw[line width = 1, photon] (0,0) -- (1,1);
    \draw[line width = 1, photon] (0,0) -- (1,-1);
    \draw[line width = 1, dashed] (-1,0) node[midway, above, xshift=-10] {} -- (0,0);
    \filldraw[fill=gray!5, line width=1.2](-1,0) circle (0.15) node {$\times$};
    \end{tikzpicture}
\end{gathered} + \begin{gathered}
    \begin{tikzpicture}[line width=1,photon/.style={decorate, decoration={snake, amplitude=1pt, segment length=6pt}}]
    \draw[line width = 1, photon] (0,0.5) -- (1,1.5);
    \draw[line width = 1, photon] (0,-0.5) -- (1,-1.5);
    \draw[line width = 1, photon] (0,-0.5) -- (0,0.5);
    \draw[line width = 1, dashed] (-1,-0.5) -- (0,-0.5);
    \draw[line width = 1, dashed] (-1,0.5) -- (0,0.5);
    \filldraw[fill=gray!5, line width=1.2](-1,-0.5) circle (0.15) node {$\times$};
    \filldraw[fill=gray!5, line width=1.2](-1,0.5) circle (0.15) node {$\times$};
    \end{tikzpicture}
\end{gathered}
+
\begin{gathered}
    \begin{tikzpicture}[line width=1,photon/.style={decorate, decoration={snake, amplitude=1pt, segment length=6pt}}]
    \draw[line width = 1, photon] (0,0.0) -- (1,1.0);
    \draw[line width = 1, photon] (0,0.0) -- (1,-1.0);
    \draw[line width = 1, dashed] (-1,-0.5) -- (0,0.0);
    \draw[line width = 1, dashed] (-1,0.5) -- (0,0.0);
    \filldraw[fill=gray!5, line width=1.2](-1,-0.5) circle (0.15) node {$\times$};
    \filldraw[fill=gray!5, line width=1.2](-1,0.5) circle (0.15) node {$\times$};
    \end{tikzpicture}
\end{gathered} 
+ \begin{gathered}
    \begin{tikzpicture}[line width=1, photon/.style={decorate, decoration={snake, amplitude=1pt, segment length=6pt}}]
        \filldraw[fill=gray!5, line width=1.2] (-0.3,-0.15) rectangle (0,0.15) 
            node[pos=0.5] {$\odot$}; 
        \draw[line width=1, photon] (0,0) -- (1,1);
        \draw[line width=1, photon] (0,0) -- (1,-1);
    \end{tikzpicture}
\end{gathered}
+O(G^\frac{5}{2})\,,
\end{equation}
\end{widetext}
where the last diagram denotes the insertion of the tidal operators (\ref{eq:Stidal}-\ref{eq:dissipact}). In 5D, the only divergent integral arises from the triangle diagram. The result from the evaluation of $ i\Delta^{\text{EFT}}$  in  dimensional regularization  $D=5+2\epsilon_5$ is provided in Eqs. \eqref{eq:efttree}, \eqref{eq:eft1-loop} and \eqref{eq:Tidal_Amp}  in the supplementary material.
Dotted with  this, the elastic phase-shift $\delta^{\text{EFT}}_\ell$, is computed from the inversion formula 
\begin{equation}\label{eq:inveform}
     \delta_{\ell}^{\text{EFT}} {=} \frac{ \omega}{4\pi} \frac{(4\pi / \omega^2)^{\frac{4{-}D}{2}}}{2 \Gamma(\frac{D-2}{2})} \int_{-1}^1 dz (1{-}z^2)^{\frac{D{-}4}{2}} P_{\ell}^{(D)}(z) \Delta^{\text{EFT}}( z)~,
\end{equation}
where $P_{\ell}^{(D)}(z)$ are the Gegenbauer function defined in Eq.~\eqref{eq:Gegenbauer}.  An analogous formula follows for the dissipative parameter $\eta^{\text{EFT}}_\ell$.

Explicit evaluation of the elastic phase-shift in the  EFT and the in  UV  shows complete agreement for generic $\ell\ge2$ harmonics (see Eqs. \eqref{eq:dUVgenl} and \eqref{eq:dEFTgenl} in the supplementary material). This corresponds to the matching between the UV far-zone and its point-particle EFT counterpart. 
At order $G^2$, $\ell =0$ and $\ell=1$  excitation receive  tidal  contributions, as expected from Eq.~\eqref{eq:tildal5D}.  Imposing the matching condition $\delta_{5,\ell=0,1 }^{\text{EFT}} = \delta_{5,\ell =0,1}^{\text{UV}}$, allows us to fix the bare free coefficients in the tidal action \eqref{eq:Stidal}:
\begin{equation}\label{eq:lovenumbers}
    \boxed{
    \begin{split}
       \frac{ c_{\omega^2 \phi,0} }{\pi^2 r_{s,5}^4 }&{=} {-}  \left(2 \log \left(\mu r_{s,5}\right)+\frac{11}{6}+\gamma_E +\log (\pi )+\frac{1}{\epsilon _5} \right)  \,,\\
       \frac{ -c_{\phi,1}}{\pi^2 r_{s,5}^4} &{=}\left(2 \log \left(\mu  r_{s,5}\right)+\frac{5}{6}+\gamma_E +\log \left(\frac{\pi }{16}\right)+\frac{1}{\epsilon _5}\right)\frac{1}{16}\,,
    \end{split}    
    }
\end{equation}
where the $\frac{1}{\epsilon_5}$ contributions regularize the   UV-divergence on the EFT side. Finite renormalized Love numbers can be obtained   by   fixing a subtraction scheme (see supplementary material).   Note also $\log\omega$ contributions cancel between the EFT and the UV results, thus providing a consistency check of the matching procedure.

From Eq.~\eqref{eq:lovenumbers}, we obtain the   beta functions 
\begin{align}
    \mu \frac{d   c_{\omega^2 \phi,0} }{d\mu} = -2 \pi ^2 r_{s,5}^4\,,\quad
   \mu \frac{d   c_{\phi,1}}{d\mu} = -\frac{1}{8} \pi ^2 r_{s,5}^4\,.
\end{align}
which recover the known results in the literature \cite{Kol:2011vg,Hui:2020xxx,Ivanov:2022qqt,Ivanov:2024sds}.

The matching to the dissipative contribution Eqs. \eqref{eq:etaUVl0} and \eqref{eq:etaEFTl0}, fixes the leading order dissipative number 
\begin{equation}
\boxed{
    c_{\omega \phi,0} =\frac{4 (A-B)}{A+B} \pi^2 r_{s,5}^3 }\,.
\end{equation}

Note that, as expected, the values for the matched elastic Love numbers in Eq.~\eqref{eq:lovenumbers} at leading PM order are independent of the type of BCs imposed at the BH horizon, but receive  near and far zone contributions from the UV computation; in the fixed $\ell$-prescription, the near-far factorization  of the UV solutions breaks down \cite{Bautista:2023sdf}.  The dissipative Love number does depend on the choice of BCs even at leading PM order.

\textit{Discussion}--
In this paper, we have studied general frequency, scalar linear perturbations of 5D STBH, providing both, a UV-complete solution for the scattering problem and, for the first time, an on-shell matching for the static  $\ell =1$ and dynamical $\ell=0$ elastic Love numbers for the STBH at 2PM, and the $\ell=0$, $G^{3/2}$ dissipative Love number, 
for a generic class of boundary conditions imposed at the BH horizon. Non-trivial near and far zone contributions to the tidal coefficients are manifest, and some universality in the leading PM elastic Love numbers is observed.  
Higher PM EFT computations proving even-$\ell$ static Love numbers are left for future work, although from the discussion around Eq.~\eqref{eq:tildal5D} we expect them to be RG-independent and vanishing.  

Although we have obtained the connection formula \eqref{eq:Bratio} using as main example massless scalar perturbations to the 5D STBH,   the formula is also valid for massless  
perturbations of spin-weight $s=1,2$. Indeed, once the perturbation equations for spin-weight-$s$ are written in terms of gauge invariant master variables \cite{Kodama:2003kk,Kodama:2007ph,Hui:2020xxx}, it is easy to show they correspond to  RCHE  with their respective gauge-gravity dictionary, even with the absence of electric-magnetic duality in 5D. Similarly,  for generic spacetime dimensions, the radial equation still has two regular singular points located at  $r=0$ and $r=r_{s,D}$, the position of the  BH-horizon in D-dimensions, but now the irregular singularity at $r=\infty$ has Poincaré rank  $h=\frac{1}{D-3}$. In $D=4,5$ one recovers the confluent, and reduced confluent Heun equations respectively, whereas in the $D\to\infty$ limit, the radial equation becomes the hypergeometric equation.

\textit{Acknowledgments.}
We would like to thank Sujay K. Ashok, Alok Laddha, Arkajyoti Manna, Akavoor Manu, Giorgio Di Russo, and Tanmoy Sengupta for useful discussions. The work of Y.F.B.   has been supported by the European Research Council under Advanced Investigator Grant ERC–AdG–885414. The work of C.I. is supported by the Swiss National Science Foundation Grant No. 185723.

\appendix


\begin{widetext}

\section{Connection formula derivation details}
In this section we provide the details for the derivation of the connection formula in Eq.~\eqref{eq:Bratio}.
\subsection{Incoming boundary conditions at the BH horizon}
 As a warm-up up we consider purely ibcs first.
Near the BH-horizon,  the differential equation \eqref{eq:rcheun} has two solutions 
\begin{equation}
   w(z) \simeq (1-z)^{\frac{1}{2}\pm a_1} \, ~~~~~\text{as} ~~~~z \rightarrow 1 \,. 
\end{equation}
In order to impose  purely ingoing wave-modes at the horizon, one has to choose for $w(z)$ 
\begin{align}
    w^{\text{ibc.}}(z) \simeq (1-z)^{\frac{1}{2}+a_1} \, ~~~~~\text{as} ~~~~z \rightarrow 1 \,.
\end{align}
This solution can now be expanded close to $z = \infty$ or equivalently $r = \infty$. We find (see Ref.~\cite{Bonelli:2022ten}):
\begin{align} \label{eq:connection5_sol}
    w^{\text{ibc.}}(z) =  D^{-}_{lm \omega} e^{L \sqrt{z}} \sqrt{z}^{-1/2} (1 + \mathcal{O}(z^{-1/2})) +  D^{+}_{lm \omega} e^{-L \sqrt{z}} \sqrt{z}^{-1/2} (1 + \mathcal{O}(z^{-1/2})) \,,
\end{align}
where $ D^{\pm}_{lm \omega}$ are the connection coefficients of the RCHE given by
\begin{align}\label{eq:coeffD}
     D^{-}_{lm \omega} = \sum_{\sigma=\pm} L^{-\frac{1}{2} + 2\sigma a} \frac{\Gamma(1-2 \sigma a) \Gamma(-2\sigma a) \Gamma(1+ 2a_1) 2^{-2\sigma a}}{\sqrt{2\pi} \, \Gamma(\frac{1}{2}+ a_1 + a_0 -\sigma a) \Gamma(\frac{1}{2}+ a_1 - a_0 -\sigma a) } e^{-\frac{\sigma}{2}\partial_a F }  \,.
    \end{align}
$ D^{+}_{lm \omega}$ can be obtained from ${}_{s} D^{-}_{lm \omega}$ via $L \rightarrow -L$.
Next, we map Eq.~\eqref{eq:connection5_sol} to the asymptotic behavior given in Eq.~\eqref{eq:radial_asym}. As $r \rightarrow \infty$, we have
\begin{align}
    z^{\frac{1}{4}}(z-1)^{-\frac{1}{2}} e^{L \sqrt{z}} \sqrt{z}^{-1/2} \simeq e^{-i \omega r} r^{-1} \left(r_{s,5}\right) \,, \cr
    z^{\frac{1}{4}}(z-1)^{-\frac{1}{2}} e^{-L \sqrt{z}} \sqrt{z}^{-1/2} \simeq e^{i \omega r} r^{-1}\left(r_{s,5}\right) \,,
\end{align}
where  we have used for the  tortoise coordinate, $r_\star = r + \frac{1}{2} r_{s,5} \log{\left\vert \frac{r- r_{s,5}}{r+ r_{s,5}} \right\vert}$, so that
\begin{align}
    r^{-3/2} e^{-i \omega r_\star} \simeq r^{-3/2} e^{-i \omega r}  \,, \cr
    r^{-3/2} e^{i \omega r_\star} \simeq r^{-3/2} e^{i \omega r}  \,.
\end{align}
Finally, the ratio of the asymptotic coefficients in Eq.~\eqref{eq:radial_asym} is
\begin{align}
    \frac{ B^{ref}_{lm}}{ B^{inc}_{lm}} = \frac{D^{+}_{lm \omega}}{D^{-}_{lm \omega}}\,,
\end{align}
which, after simplifying, becomes
 \begin{align}\label{eq:bratappprev}
     \frac{ B^{ref}_{lm}}{ B^{inc}_{lm}} &= e^{-i \frac{\pi}{2}}  \frac{\sum_{\sigma=\pm}  \frac{\Gamma(1-2 \sigma a) \Gamma(-2\sigma a) (-L)^{2\sigma a} 2^{-2\sigma a}e^{-\frac{\sigma}{2}\partial_a F}}{\Gamma(\frac{1}{2}+ a_1 + a_0 -\sigma a) \Gamma(\frac{1}{2}+ a_1 - a_0 -\sigma a) } }{\sum_{\sigma'=\pm}  \frac{\Gamma(1-2 \sigma' a) \Gamma(-2\sigma' a) L^{2\sigma' a} 2^{-2\sigma'a } e^{-\frac{\sigma'}{2}\partial_a F}}{\Gamma(\frac{1}{2}+ a_1 + a_0 -\sigma' a) \Gamma(\frac{1}{2}+ a_1 - a_0 -\sigma' a)}} \,.
 \end{align}
A  further simplification turns this into
\begin{align}\label{eq:bratapp}
     \frac{ B^{ref}_{lm}}{ B^{inc}_{lm}} &= e^{i\pi (2a -\frac{1}{2})} \times \frac{1 + e^{-2i\pi a} \mathcal{K}_5}{1 + e^{2i\pi a}\mathcal{K}_5}\,,
\end{align}
where we have used $-1 = e^{i\pi}$ and $-i = e^{-i\pi/2}$, and the  tidal response function, $\mathcal{K}_5$ defined in  Eq.~\eqref{eq:response5d}. Using Eq.~\eqref{eq:bratapp} into   Eq.~\eqref{eq:deffdelta}, we recover the result given in Eq.~\eqref{eq:Bratio} in the main text with $B=0$.

\subsection{Outgoing boundary conditions at the BH horizon}
In a similar way, we can instead impose purely outgoing wave modes at the BH horizon. This is, take
\begin{align}\label{eq:outgoing_sol}
    w^{\text{obc.}}(z) \simeq (1-z)^{\frac{1}{2}-a_1} \, ~~~~~\text{as} ~~~~z \rightarrow 1 \,.
\end{align}
Now one can expand the equation close to $z = \infty$ or $r = \infty$. The connection formula for the outgoing solution can be obtained via the replacement  $a_1 \rightarrow -a_1$ in the ingoing one \cite{Bonelli:2022ten}. We simply arrive at the ratio of asymptotic coefficients
\begin{align}\label{eq:bratobc}
     \frac{ B^{ref}_{lm}}{ B^{inc}_{lm}}\Big|_{\text{obc.}} &= e^{i\pi (2a -\frac{1}{2})} \times \frac{1 + e^{-2i\pi a} \mathcal{K}_{5}^{\text{obc.}}}{1 + e^{2i\pi a}\mathcal{K}_{5}^{\text{obc.}}} \,.
\end{align}
Here $\mathcal{K}_{5}^{\text{obc.}}$ is the tidal response function for the outgoing solution given by 
\begin{align}
    \mathcal{K}_{5}^{\text{obc.}} &= \mathcal{K}_{5} |_{a_1 \rightarrow-a_1} \cr
    &=\vert L\vert^{-4a} 2^{4a} \frac{\Gamma(2a) \Gamma(1+2a)  \Gamma(\frac{1}{2}- a_1 + a_0 - a) \Gamma(\frac{1}{2}- a_1 - a_0 - a)}{\Gamma(-2a) \Gamma(1-2a)  \Gamma(\frac{1}{2}- a_1 + a_0 + a) \Gamma(\frac{1}{2}- a_1 - a_0 + a)} e^{\partial_a F} \,.
\end{align} 
recovering the result in the main text \eqref{eq:Bratio} for $A=0$.

\subsection{Semi-Reflective boundary conditions at the BH horizon}
A more general class of semi-reflecting boundary conditions can be imposed at the BHs. We take for the superposition 
\begin{align}
    w^{\text{srbc.}}(z) \simeq A (1-z)^{\frac{1}{2}+a_1} +B(1-z)^{\frac{1}{2}-a_1}  \, ~~~~~\text{as} ~~~~z \rightarrow 1 \,.
\end{align}
where $A,B$ are real numbers. The results for ibc. (obc.) are  recovered by setting $B=0\,,\, A=1$ ($A=0\,,\,B=1$).

For this generic choice of boundary conditions,  the ratio of coefficients  is 
\begin{align}
 \frac{ B^{ref}_{lm}}{ B^{inc}_{lm}}\Big|_{\text{srbc.}}  &= \frac{A  \, D^{+}_{lm \omega}(a_1) + B \, D^{+}_{lm \omega}(-a_1)}{A \,  D^{-}_{lm \omega}(a_1) + B \, D^{-}_{lm \omega}(-a_1)} \cr
     &= \frac{D^{+}_{lm \omega}(a_1)}{D^{-}_{lm \omega}(a_1)} \frac{\left(1 + \frac{B}{A} \frac{D^{+}_{lm \omega}(-a_1)}{D^{+}_{lm \omega}(a_1)}\right)}{\left(1 + \frac{B}{A} \frac{D^{-}_{lm \omega}(-a_1)}{D^{-}_{lm \omega}(a_1)}\right)} \cr
     &= e^{i\pi (2a -\frac{1}{2})} \times \frac{1 + e^{-2i\pi a} \mathcal{K}_{5}}{1 + e^{2i\pi a}\mathcal{K}_{5}} \left[ \frac{1 +  \mathcal{R}\frac{B}{A}\left(\frac{1 + e^{-2i\pi a} \mathcal{K}_{5}|_{a_1 \rightarrow-a_1}}{1 + e^{-2i\pi a} \mathcal{K}_{5}} \right)}{1 +  \mathcal{R}\frac{B}{A}\left(\frac{1 + e^{2i\pi a} \mathcal{K}_{5}|_{a_1 \rightarrow-a_1}}{1 + e^{2i\pi a} \mathcal{K}_{5}} \right) }  \right]\,,
\end{align}
where
\begin{align}
\label{eq:Rgammas}
    \mathcal{R} = \frac{\Gamma(1-2a_1)}{\Gamma(1+2a_1)} \frac{\Gamma(\frac{1}{2}+a_0 +a _1 -a)\Gamma(\frac{1}{2}-a_0 +a _1 -a)}{\Gamma(\frac{1}{2}-a_0 -a _1 -a)\Gamma(\frac{1}{2}+a_0 -a _1 -a)} \,.
\end{align}

 Replacing these results  into Eq.~\eqref{eq:deffdelta} recovers Eq.~\eqref{eq:Bratio}.
Finally, at low energies $a$ remains real and therefore the phase-shift can be evaluated via
\begin{equation}
    2i\delta_{5,\ell} ={\red{i\pi(2a+\ell+1)}} + i \text{Arg}\left[{\blue{\frac{1 + e^{-2i\pi a} \mathcal{K}_5}{1 + e^{2i\pi a}\mathcal{K}_5}
\left[ \frac{1 +  \mathcal{R}\frac{B}{A}\left(\frac{1 + e^{-2i\pi a} \mathcal{K}_{5}|_{a_1 \rightarrow-a_1}}{1 + e^{-2i\pi a} \mathcal{K}_{5}} \right)}{1 +  \mathcal{R}\frac{B}{A}\left(\frac{1 + e^{2i\pi a} \mathcal{K}_{5}|_{a_1 \rightarrow-a_1}}{1 + e^{2i\pi a} \mathcal{K}_{5}} \right) }  \right]}}\right]\,,
\end{equation}
and similarly,  the absorption parameter is obtained from  
\begin{equation}
   \eta_{5,\ell} = \left|{\blue{\frac{1 + e^{-2i\pi a} \mathcal{K}_5}{1 + e^{2i\pi a}\mathcal{K}_5}
\left[ \frac{1 +  \mathcal{R}\frac{B}{A}\left(\frac{1 + e^{-2i\pi a} \mathcal{K}_{5}|_{a_1 \rightarrow-a_1}}{1 + e^{-2i\pi a} \mathcal{K}_{5}} \right)}{1 +  \mathcal{R}\frac{B}{A}\left(\frac{1 + e^{2i\pi a} \mathcal{K}_{5}|_{a_1 \rightarrow-a_1}}{1 + e^{2i\pi a} \mathcal{K}_{5}} \right) }  \right]}}\right|\,.
\end{equation}
\section{The Nekrasov-Shatashvili  function for the RCHE and explicit UV phase-shift results }
\subsection{The NS function}\label{sec:NSfpr}
Let us denote a Young tableau as $Y = (\nu_1' \geq \nu_2' \geq \ldots)$ and its transpose as $Y^T = (\nu_1 \geq \nu_2 \geq \ldots)$. Accordingly, we define the so-called leg and arm length, respectively $L_Y$ and $A_Y$, as
\begin{equation}
A_Y(i,j) = \nu_i' - j, \quad L_Y(i,j) = \nu_j - i \,.
\end{equation}
Then the NS function for the reduced Confluent Heun problem  is given by
\begin{equation}\label{eq:NS}
F(a_0,a_1,L) = \lim_{b \to 0} b^2 \log \sum_{\vec{Y}} \left( \frac{L^2}{4 b^2} \right)^{|\vec{Y}|} 
z_{\text{vec}}(a/b, \vec{Y}) 
\prod_{\theta = \pm} z_{\text{hyp}}(a/b, \vec{Y}, a_1/b + \theta a_0/b) \, .
\end{equation}
where $a_0,a_1$ and $L$ are the parameters of the RCHE \eqref{eq:rcheun}. Here $\vec{Y} = (Y_1, Y_2)$ denotes a pair of Young tableaux and
\begin{equation}
\begin{aligned}
&z_{\text{hyp}}(\alpha, \vec{Y}, \mu) = 
\prod_{k=1,2} \prod_{(i,j) \in Y_k} 
\left( \alpha_k + \mu + b^{-1} \left( i - \tfrac{1}{2} \right) + b \left( j - \tfrac{1}{2} \right) \right), \\
&z_{\text{vec}}(\alpha, \vec{Y}) = 
\prod_{k,l=1,2} \prod_{(i,j) \in Y_k} 
E^{-1}(\alpha_k - \alpha_l, Y_k, Y_l, (i,j)) 
\prod_{(i',j') \in Y_l} 
\left( Q - E(\alpha_l - \alpha_k, Y_l, Y_k, (i',j')) \right)^{-1}, \\
&E(\alpha, Y_1, Y_2, (i,j)) = 
\alpha - b^{-1} L_{Y_2}((i,j)) + b \left( A_{Y_1}((i,j)) + 1 \right) \,, \\
\end{aligned}
\end{equation}
with
\begin{equation}
\alpha_k = \begin{cases} \alpha,\,\,\, k=1 \,,  \\ -\alpha, \,\,\, k=2 \,. \end{cases}
\end{equation}
For instance, evaluation of the NS function in \eqref{eq:NS}  up to order $L^6$ produces
\begin{equation}\label{eq:FL6}
\begin{split}
    F =& \frac{L^2 \left(-4 a^2+4 a_0^2-4 a_1^2+1\right)}{32 a^2-8}+\frac{L^4}{1024 \left(a^2-1\right) \left(4 a^2-1\right)^3} \left(-64 a^6+48 a^4 \left(8 a_0^2+8 a_1^2+1\right)-4 a^2 \left(80 a_0^4\right.\right.\\
    &\left.\left.+a_0^2 \left(48-160 a_1^2\right)+80 a_1^4+48 a_1^2+3\right)-112 \left(a_0^2-a_1^2\right)^2+24 a_0^2+24 a_1^2+1\right)\\
    &+\frac{L^6 (a_0-a_1) (a_0+a_1)}{768 \left(4 a^2-1\right)^5 \left(4 a^4-13 a^2+9\right)} \left(16 \left(80 a^8-16 a^6 \left(14 a_0^2+14 a_1^2+5\right)+6 a^4 \left(24 a_0^4-4 a_0^2 \left(12 a_1^2+1\right)\right.\right.\right.\\
    &\left.\left.\left.+24 a_1^4-4 a_1^2+5\right)+a^2 \left(232 (a_0^4+a_1^4)+a_0^2 \left(54-464 a_1^2\right)+54 a_1^2-5\right)+29 \left(a_0^2-a_1^2\right)^2\right)-136 (a_0^2+ a_1^2)+5\right)+ \mathcal{O}(L^8)    
\end{split}
\end{equation}

Let us also provide for illustrative purposes the explicit solution of the Matone relation \eqref{eq:matone} up to order $L^6$. 
\begin{equation}\label{eq:aL6}
    \begin{split}
        a=& -\frac{1}{2} \sqrt{1-4 u}+
        \frac{L^2 \left(a_0^2-a_1^2+u\right)}{8 \sqrt{1-4 u} u}-\frac{L^4} {128 (1-4 u)^{3/2} u^3 (4 u+3)}\left(a_0^4 (3-5 u (12 u+1))+2 a_0^2 \left(a_1^2 \left(60 u^2+5 u-3\right)\right.\right.\\
        &\left.\left.+u (3-u (36 u+11))\right)+a_1^4 (3-5 u (12 u+1))+2 a_1^2 u (u (12 u+17)-3)-u^3 (12 u+5)\right) \\
        &+\frac{L^6 }{1024 (1-4 u)^{5/2} u^5 (u+2) (4 u+3)^2}\left(a_0^6 (u (10 u (7 u (24 u (2 u+3)+7)-43)-51)+18)\right.\\
        &\left.+a_0^4 \left(u (u (2 u (7 u (40 u (10 u+19)+181)-459)-225)+54)-3 a_1^2 (u (10 u (7 u (24 u (2 u+3)+7)-43)-51)+18)\right)\right.\\
        &\left.+a_0^2 \left(3 a_1^4 (u (10 u (7 u (24 u (2 u+3)+7)-43)-51)+18)-6 a_1^2 u (u+2) (2 u (7 u (40 u (2 u+1)-7)-24)+9)\right.\right.\\
        &\left.\left.+2 u^2 (u (u (u (40 u (30 u+67)+1039)-110)-102)+18)\right)-a_1^6 (u (10 u (7 u (24 u (2 u+3)+7)-43)-51)+18)\right.\\
        &\left.+a_1^4 u (u (14 u (u (40 u (2 u+11)+257)-39)-297)+54)-2 a_1^2 u^2 (u (u (u (40 u (6 u+19)+1171)+136)-138)+18)\right.\\
        &\left.+2 u^5 (u+2) (40 u (2 u+1)+17)\right)
        + \mathcal{O}(L^8)\,.
    \end{split}
\end{equation}

\subsection{UV phase-shift and absorption factor}

Using these expressions for  $F$  and $a$ in the connection formula  Eq.~\eqref{eq:Bratio},  and with the aid of the dictionary in Eq.~\eqref{eq:dictionary},  we obtain the explicit PM-expanded solution for the UV phase-shift for generic boundary conditions 
\begin{align}
\delta_{5,\ell }^{\text{UV}} =&{\red{\frac{3 \pi  (\omega  r_{s,5})^2}{8 (\ell +1)} + \frac{\pi  (5 \ell  (\ell +2) (7 \ell  (\ell +2)-17)-48) (\omega  r_{s,5})^4}{128 (\ell -1) \ell  (\ell +1)^3 (\ell +2) (\ell +3)}}}\nonumber\\
&{\red{+\frac{3 \pi  \omega ^6 \left(7 (\ell -1) \ell  (\ell +2) (\ell +3) (11 \ell  (\ell +2)-24) (\ell +1)^2+288\right) r_{s,5}^6}{512 \ell ^2 (\ell +1)^5 \left(\ell ^3+4 \ell ^2+\ell -6\right)^2}}}  
+  \mathcal{O}(\omega  r_{s,5})^8\,,\quad \ell \ge2 \,,\label{eq:dUVgenl}\\
\delta_{5,\ell=0 }^{\text{UV}}   = &{\red{\frac{3}{8} \pi  \omega ^2 r_{s,5}^2+\frac{17}{384} \pi  \omega ^4 r_{s,5}^4}}
 {\blue{-\frac{1}{64} \pi  \omega ^4 r_{s,5}^4 \left(8 \log \left(\omega  r_{s,5}\right)+8 \gamma_E -5-\log (256)\right)}}
 +  \mathcal{O}(\omega  r_{s,5})^6\,,\label{eq:dUVl0}\\
  \delta_{5,\ell=1 }^{\text{UV}}    = & {\red{\frac{3}{16} \pi  \omega ^2 r_{s,5}^2+\frac{463 \pi  \omega ^4 r_{s,5}^4}{12288}}} {\blue{-\frac{\pi  \omega ^4 r_{s,5}^4 \left(2 \log \left(\omega  r_{s,5}\right)+2 \gamma_E -1-\log (64)\right)}{1024}}} + \mathcal{O}(\omega  r_{s,5})^6\,.\label{eq:dUVl1}
\end{align} 
and similarly the absorption factor for $\ell=0$ 
\begin{equation}\label{eq:etaUVl0}
  \eta_{5,\ell=0 }^{\text{UV}} =
{\blue{1+ \frac{\pi  r_{s,5}^3 \omega ^3 (A-B)}{4 (A+B)}}} +  \mathcal{O}(\omega  r_{s,5})^5\,.
\end{equation}

Although for the elastic phase-shift the parameters $A,B$ controlling the boundary conditions do not appear up to the PM-orders considered here, they will appear at subleading orders, for instance, at $\mathcal{O}(\omega r_{s,5})^6$, the $\ell=1$ phase shift receives the contributions

\begin{equation}
\begin{split}
    \delta_{5,\ell=1}|_{\mathcal{O}(r_{s,5}\omega)^6}& =\blue{\frac{\pi ^3 A B \omega ^6 (\log (16)-4) r_{s,5}^6}{1024 (A+B)^2} } {\red{+\frac{1767 \pi  \omega ^6 r_{s,5}^6}{131072}}} +\\
   & {\blue{-\frac{\pi  \omega ^6 r_{s,5}^6}{16384} \left(4 \log \left(\omega  r_{s,5}\right) \left(-3 \log \left(\omega  r_{s,5}\right)-6 \gamma_E +19+18 \log (2)\right)+56 \zeta (3)-12 \gamma_E ^2+\pi ^2\right.}}\\
   &{\blue{\left.-99+\gamma_E  (76+72 \log (2))-12 \log (2) (19+\log (512))\right)}} \,.
    \end{split}
\end{equation}
Notice that although the first term vanishes for purely ibc. or purely obc., in general it does not vanish, as it is the case of, for instance,   rbc. $A=B$. The absorption factor 
\begin{equation}
  \eta_{5,\ell=1 }^{\text{UV}} =
{\blue{1+\frac{\pi ^3 \omega ^5 (A-B) r_{s,5}^5}{1024 (A+B)}}} + \mathcal{O}(\omega  r_{s,5})^7\,.
\end{equation}
Let us for completeness include also the contributions in the $\ell=2$ mode up to order $\mathcal{O}(\omega r_{s,5})^6$

\begin{equation}
  \delta_{5,\ell=2} = \red{\frac{187 \pi  \omega ^6 r_{s,5}^6}{76800}+\frac{7}{640} \pi  \omega ^4 r_{s,5}^4+\frac{1}{8} \pi  \omega ^2 r_{s,5}^2}  + \mathcal{O}(\omega  r_{s,5})^8
\end{equation}
which, as expected, does not have any near-zone traces up to this PM order.

\subsection{Large Frequency Behavior of the NS Function}
The rank-$1/2$ irregular conformal block can be estimated via the following irregular state \cite{Bonelli:2022ten}
\begin{equation}
    \begin{aligned}
& \left\langle\Lambda^2\right| L_0=\Lambda^2 \partial_{\Lambda^2}\left\langle\Lambda^2\right| \\
& \left\langle\Lambda^2\right| L_{-1}=-\frac{\Lambda^2}{4}\left\langle\Lambda^2\right| \\
& \left\langle\Lambda^2\right| L_{-n}=0, \quad n>1 ~,
\end{aligned}
\end{equation}
where $L_{n}, n\in \mathbb{Z}$ are the Virasoro generators. The NS function can be defined as the following classical conformal block
\begin{equation}
    F = \lim_{b^2 \rightarrow 0} b^2 \log \langle \Lambda^2| V_{\alpha_1}(1)|\Delta_{\alpha_0}\rangle_{\rm CB} ~,
\end{equation}
where the subscript CB indicates that we only consider the conformal block associated with the correlators.
We want to evaluate the above expression in the limit where $\alpha_1 \gg \alpha_0$. Note that the shift of the operators is given by
\begin{equation}
    V_{\alpha_1}(1) = e^{L_{-1}} V_{\alpha_1}(0) e^{-L_{-1}} ~.
\end{equation}
and therefore
\begin{equation}
    \langle \Lambda^2| V_{\alpha_1}(1) | \Delta_{\alpha_0}\rangle = \langle \Lambda^2| e^{L_{-1}} V_{\alpha_1}(0) e^{-L_{-1}} |\Delta_{\alpha_0}\rangle = e^{-\Lambda^2 /4}\langle \Lambda^2| V_{\alpha_1}(0) e^{-L_{-1}}|\Delta_{\alpha_0}\rangle 
\end{equation}
In the limit, $\alpha_1 \gg \alpha_0$, we can make the following simplification by treating $|\Delta_{\alpha_0}\rangle$ approximately as vacuum $|0 \rangle$. Since the CFT vacuum is invariant under translation, i.e. $e^{-L_{-1}}|0\rangle = |0 \rangle$ and thus
\begin{equation}
    \langle \Lambda^2| V_{\alpha_1}(1) | \Delta_{\alpha_0}\rangle \simeq e^{-\Lambda^2/4} \langle \Lambda^2| \Delta_{\alpha_1}\rangle ~, \quad \alpha_1 \gg \alpha_0 ~.
\end{equation}
Then, we get the 
\begin{equation}
    F \simeq -\frac{L^2}{4} ~, \quad a_1 \gg a_0 ~.
\end{equation}
Combining with the dictionary given in Eq.~\eqref{eq:dictionary}, we arrive at 
\begin{equation}
    a^2 \simeq \frac{L^2}{4} = - \frac{1}{4} (r_{s,5}\omega)^2 ~, \quad |r_{s,5} \omega | \gg 1~.
\end{equation}
According to Eq.~\eqref{eq:bratappprev}, the connection formula is invariant under $a \rightarrow -a$, so we can choose either branch. In this paper, we choose 
\begin{equation}
    a \simeq - \frac{i}{2} (r_{s,5}\omega) ~, \quad |r_{s,5} \omega | \gg 1 ~.
\end{equation}

\section{Gravitational Raman Scatterings through $\mathcal{O}(G^2)$}

\subsection{Partial Wave Transformation and Exponential Representation of S-matrix}
In this appendix, we provide details on the calculation of the one-loop wave scattering amplitude off the black holes in general $D$ dimensions. We then specialize to $D=5$ for comparison to the UV solutions presented in the main text.  The standard parametrization of the S-matrix is given by the following expansion 
\begin{equation}
    \boldsymbol{S} \equiv \mathbf{1} + i \boldsymbol{T} ~,
\end{equation}
where $\boldsymbol{T}$ is the scattering operator.  Considering the following 1-to-1 scattering amplitude within the background gravitational source
\begin{equation}
    \left\langle k_2\right| i \boldsymbol{T} \left|k_1\right\rangle=i \mathcal{M}\left(k_1 \rightarrow k_2\right) \cdot(2 \pi) \delta\left(u \cdot\left(k_1+k_2\right)\right) ~,
\end{equation}
where $k_1, k_2$ are the four momenta for the incoming and outgoing particles, and $u$ is the four momentum of the black hole. The $\delta$-function above makes manifest energy conservation due to time translation invariance along the BH worldline. The direction of $k_i$ transverse to $u$ is denoted as $n_i$. The partial wave expansion of the momentum space amplitude can be derived by projecting the amplitude $\mathcal{M}$ onto the generalized spherical harmonics
\begin{equation}
    \left(\eta_{\ell} e^{2 i \delta_{\ell}}-1\right)=\frac{i \omega}{2 \pi} \frac{\left(4 \pi / \omega^2\right)^{\frac{4D}{2}}}{2 \Gamma\left(\frac{D-2}{2}\right)} \int_{-1}^1 d z\left(1-z^2\right)^{\frac{D-4}{2}} P_{\ell}^{(D)}(z) \mathcal{M}(\omega, z) ~,
\end{equation}
with the Gegenbauer function
\begin{equation}\label{eq:Gegenbauer}
    P_{\ell}^{(D)}(z)={ }_2 F_1\left(-\ell, \ell+D-3, \frac{D-2}{2}, \frac{1-z}{2}\right) ~, \quad z = n_1 \cdot n_2 = \cos\theta ~.
\end{equation}
The unitarity of the S-matrix 
\begin{equation}
    \boldsymbol{S} \boldsymbol{S}^\dagger = \mathbf{1}
\end{equation}
also implies another useful parametrization, known as exponential parametrization \cite{Damgaard:2021ipf}
\begin{equation}\label{eq:exp_rep}
    \boldsymbol{S} \equiv \exp(i \boldsymbol{\Delta}) ~,
\end{equation}
with
\begin{equation}
    \langle k_2, h_2| i \boldsymbol{\Delta} |k_1, h_1 \rangle = i \Delta (k_1,h_1 \rightarrow k_2,h_2) (2\pi) \delta(u \cdot (k_1 + k_2)) ~.
\end{equation}
In the calculation of Raman scattering amplitude, the scattering matrix can be obtained perturbatively in terms of Newton's constant $G$. Formally, we can write the perturbative series as
\begin{equation}
\begin{aligned}
    \boldsymbol{\Delta} & = G \boldsymbol{\Delta}_G + G^2 \boldsymbol{\Delta}_{G^2} + G^3 \boldsymbol{\Delta}_{G^3} + G^4 \boldsymbol{\Delta}_{G^4}  \\
    \boldsymbol{T} & = G \boldsymbol{T}_G + G^2 \boldsymbol{T}_{G^2} + G^3 \boldsymbol{T}_{G^3} + G^4 \boldsymbol{T}_{G^4} ~,
\end{aligned}
\end{equation}
Once plugging the above perturbative expression into the exponential parameterization in Eq.~\eqref{eq:exp_rep}, we get the relation between the $\boldsymbol{\Delta}$ and $\boldsymbol{T}$ matrix
\begin{equation}
\begin{aligned}
\label{eq:reverseU1}
    \Delta_G & = T_G ~, \\
    \Delta_{G^2} & = T_{G^2} - \frac{i}{2} T_{G} T_{G}~,  \\
    \cdots & \cdots
\end{aligned}
\end{equation}
This expression can be  made even simpler by making use of the perturbative unitarity condition
\begin{equation}
    T-T^{\dagger}=i T T^{\dagger}  \Rightarrow 2 {\rm Im} T_{G^2} = T_{G} T^\dagger_{G} = T_{G} T_{G} ~,
\end{equation}
which naturally generates the following simple formulas
\begin{equation}
        \begin{aligned}
            i \Delta_G & = i T_{G} ~, \\
            i  \Delta_{G^2} & = {\rm Im} (i T_{G^2}) ~.
        \end{aligned}
\end{equation}
To get the partial wave phase-shift $\delta_\ell$ we can replace explicit solutions for $i \Delta$ in the inversion formula \eqref{eq:inveform}.

\subsection{Background Field Method and Amplitudes up to  $\mathcal{O}(G^2)$}
For explicit evaluation of    $i \Delta$, we construct the integrand using the background field method. For scalar perturbations in general $D$ dimensions, we  study
\begin{equation}
    S = -\frac{1}{2} \int d^D x \sqrt{-\bar g} \bar g^{\mu\nu}\partial_\mu \phi \partial_\nu \phi =- \frac{1}{2} \int d^D x \eta^{\mu\nu} \partial_\mu \phi \partial_\nu \phi - \frac{1}{2} \int d^D x (\sqrt{-\bar g} \bar g^{\mu\nu} - \eta^{\mu\nu}) \partial_\mu \phi \partial_\nu \phi ~.
\end{equation}
The first term on the right-hand side of the last equality can be thought of as free fields on the flat background, and the second term captures the interactions between scalar fields and the off-shell gravitons. It turns out to be easier to compute scattering amplitudes in isotropic coordinates where the spherically symmetric D-dimensional BH metric is 
\begin{equation}
    g_{\mu\nu} = \left(1+\frac{\mu}{4|\boldsymbol{r}|^{D-3}}\right)^{4 /(D-3)}\left(\eta_{\mu \nu}+u_\mu u_\nu\right)-\left(\frac{1-\frac{\mu}{4 | \boldsymbol{r}|^{D-3}}}{1+\frac{\mu}{4|\boldsymbol{r}|^{D-3}}}\right)^2 u_\mu u_\nu ~,
\end{equation}
where $u^\mu$ is the four velocity of the BH and $\mu \equiv  \frac{16 \pi G M}{(D-2) \Omega_{D-2}}$, and $\Omega_{D-2}$ is the volume of $D-2$ dimension sphere. In 5D, one can recover the STBH metric in Eq.~\eqref{eq:metric} via a coordinate transformation.

Derivation of Feynman rules is straightforward. These are then used to compute the scattering phase $i\Delta$ order by order in $G$. At the tree level, we get
\begin{equation}\label{eq:efttree}
    i \Delta_{G} = \begin{gathered}
    \begin{tikzpicture}[line width=1,photon/.style={decorate, decoration={snake, amplitude=1pt, segment length=6pt}}]
    \draw[line width = 1, photon] (0,0) -- (1.3,1);
    \draw[line width = 1, photon] (0,0) -- (1.3,-1);
    \draw[line width = 1, dashed] (-1.2,0) node[midway, above, xshift=-10] {\footnotesize $G M /r^{D-3}$} -- (0,0);
    \filldraw[fill=gray!5, line width=1.2](-1.2,0) circle (0.15) node {$\times$};
    \end{tikzpicture}
\end{gathered}
= i \frac{4 G M \pi}{x^2} ~, \quad x^2 \equiv \sin^2 \Big(\frac{\theta}{2}\Big) = \frac{1-z}{2} ~.
\end{equation}
The wavy lines here represent the scalar fields, and dashed lines are for off-shell gravitons. Here we have used the parametrization for the massless momenta $k_1$  and $k_2$ such that $u\cdot k_1= u\cdot k_2=0$ and $k_1\cdot k_2 = -4\omega^2 \sin^2(\theta/2) $.

At one-loop, we get
\begin{equation}
\begin{aligned}
    i \Delta_{G^2} & = \begin{gathered}
    \begin{tikzpicture}[line width=1,photon/.style={decorate, decoration={snake, amplitude=1pt, segment length=6pt}}]
    \draw[line width = 1, photon] (0,0.5) -- (1,1.5);
    \draw[line width = 1, photon] (0,-0.5) -- (1,-1.5);
    \draw[line width = 1, photon] (0,-0.5) -- (0,0.5);
    \draw[line width = 1, dashed] (-1,-0.5) -- (0,-0.5);
    \draw[line width = 1, dashed] (-1,0.5) -- (0,0.5);
    \filldraw[fill=gray!5, line width=1.2](-1,-0.5) circle (0.15) node {$\times$};
    \filldraw[fill=gray!5, line width=1.2](-1,0.5) circle (0.15) node {$\times$};
    \end{tikzpicture}
\end{gathered}
+
\begin{gathered}
    \begin{tikzpicture}[line width=1,photon/.style={decorate, decoration={snake, amplitude=1pt, segment length=6pt}}]
    \draw[line width = 1, photon] (0,0.0) -- (1,1.0);
    \draw[line width = 1, photon] (0,0.0) -- (1,-1.0);
    \draw[line width = 1, dashed] (-1,-0.5) -- (0,0.0);
    \draw[line width = 1, dashed] (-1,0.5) -- (0,0.0);
    \filldraw[fill=gray!5, line width=1.2](-1,-0.5) circle (0.15) node {$\times$};
    \filldraw[fill=gray!5, line width=1.2](-1,0.5) circle (0.15) node {$\times$};
    \end{tikzpicture}
\end{gathered}
\\
& = \frac{-2 i (G M \omega)^2}{x \sqrt{1-x^2}} \Im\Bigg[4 \text{Li}_2\left(-\frac{i x}{\sqrt{1-x^2}}\right)-4 \text{Li}_2\left(\frac{i x}{\sqrt{1-x^2}}\right)-2 \text{Li}_2\left(\frac{1}{2}-\frac{i
   x}{2 \sqrt{1-x^2}}\right)+2 \text{Li}_2\left(\frac{i x}{2 \sqrt{1-x^2}}+\frac{1}{2}\right) \\
   & \quad -\log ^2\left(-1+\frac{i x}{\sqrt{1-x^2}}\right)+\log ^2\left(1+\frac{i
   x}{\sqrt{1-x^2}}\right)+2 i \pi  \log \left(-1+\frac{i x}{\sqrt{1-x^2}}\right)+\frac{-4 \pi +4 i \log (2x^2)}{\sin x} +3 \pi ^2\Bigg] \\
   & \quad + i(G M \omega)^2 \Big[x^2 \left(\frac{8}{9} \log \Big(\frac{x^2 \omega^2}{\pi \mu^2}  e^{\gamma_E} \Big) -\frac{32}{27}-\frac{8}{9 \epsilon}\right)+\frac{68}{9 \epsilon}-\frac{68}{9} 
   \log \Big(\frac{x^2 \omega^2}{\pi \mu^2} e^{\gamma_E}\Big) -\frac{404}{27} \Big]\label{eq:eft1-loop}
\end{aligned}
\end{equation}
where we have specialized the result to $D=5+2\epsilon_5$.  The divergence in $\epsilon_5$ is regularized by introducing the static Love number for the  $\ell=1$ sector and dynamical Love number for the  $\ell=0$ sector as in Eq.~\eqref{eq:Stidal}. These tidal parameters serve as contact terms in the scattering amplitude
\begin{equation}
\label{eq:Tidal_Amp}
i \Delta_{\rm tidal} = 
\begin{gathered}
    \begin{tikzpicture}[line width=1, photon/.style={decorate, decoration={snake, amplitude=1pt, segment length=6pt}}]
        \filldraw[fill=gray!5, line width=1.2] (-0.3,-0.15) rectangle (0,0.15) 
            node[pos=0.5] {$\odot$}; 
        \draw[line width=1, photon] (0,0) -- (1,1);
        \draw[line width=1, photon] (0,0) -- (1,-1);
    \end{tikzpicture}
\end{gathered}
= i \omega^2 (c_{\phi,1} +c_{\omega^2\phi,0}) - 2i c_{\phi,1} \omega^2 x^2 + \omega c_{\omega \phi,0}  ~,
\end{equation}
where in the last term we have added the contribution from the dissipative action \eqref{eq:dissipact}.
Summing this to Eqs.~\eqref{eq:efttree}, \eqref{eq:eft1-loop} and ~\eqref{eq:Tidal_Amp}, we get the explicit results for $i\Delta$ in the EFT up to one-loop, which can then be used in  Eq.~\eqref{eq:EFT_Amp} for the evaluation of the EFT phase-shift.

\subsection{EFT phase-shift}
 The last task is to compute the EFT phase-shift using Eq.~\eqref{eq:EFT_Amp}. We obtain the explicit  results 
\begin{align}
    \delta_{5,\ell }^{\text{EFT}} =& \frac{3 \pi  (\omega  r_{s,5})^2}{8 (\ell +1)} + \Big[\frac{9 }{64 (\ell +1)^3} +\frac{ (17 \ell  (\ell +2)-48) }{128 (\ell -1) \ell  (\ell +1) (\ell +2) (\ell +3)} \Big]\pi  (\omega  r_{s,5})^4 +  \mathcal{O}(\omega  r_{s,5})^6\,,\quad \ell \ge2\label{eq:dEFTgenl}\\
    \delta_{5,\ell=0 }^{\text{EFT}}   = & \frac{3}{8} \pi  \omega ^2 r_{s,5}^2+\frac{\omega ^4 \left(24 c_{\omega ^2 \phi ,0}+\pi ^2 r_{s,5}^4 (48 \log (\frac{\mu}{\omega} )-24 \gamma +91+24 \log (4 \pi ))\right)}{384 \pi }+\frac{\pi  \omega ^4 r_{s,5}^4}{16 \epsilon _5}+  \mathcal{O}(\omega  r_{s,5})^6\\
  \delta_{5,\ell=1 }^{\text{EFT}}   =  & \frac{3}{16} \pi  \omega ^2 r_{s,5}^2+\frac{\omega ^4 \left(192 c_{\phi ,1}+\pi ^2 r_{s,5}^4 (24 \log (\frac{\mu}{\omega} )-12 \gamma +485+12 \log (4 \pi ))\right)}{12288 \pi }+\frac{\pi  \omega ^4 r_{s,5}^4}{1024 \epsilon _5}+ \mathcal{O}(\omega  r_{s,5})^6\,.
\end{align}

In the first line, we have included respectively the contributions from the tree level, one-loop box, and one-loop triangle diagrams to the phase-shift, and is to be evaluated for generic $\ell\ge2$. 
For the  $\ell=0,1$ lines, we have included the tree-level, the tidal contribution, the UV-divergence part of the one-loop triangle diagram regularized via $\epsilon_5$, and finite contributions coming from both the one-loop box and one-loop triangle. 

The dissipative EFT factor for $\ell=0$ is 
\begin{equation}\label{eq:etaEFTl0}
    \eta_{\ell=0}^{\text{EFT}} =1+ \frac{c_{\omega \phi,0}\omega ^3}{16 \pi } + \mathcal{O}(\omega^5) \,.
\end{equation}

\subsection{Renormalized Love Numbers}
In Eq.~\eqref{eq:lovenumbers} we have fixed the bare Love numbers by matching the UV results with the bare EFT computation. It is more physical to   define  finite renormalized Love numbers $\bar{c}_\phi$ and $\bar{c}_{\omega^2\phi ,0}$ in the $\overline{\rm MS}$ scheme. For this, we introduce the counterterm action absorbing infinite contributions from the 1-loop  computation as
\begin{equation}
    S^{\text{ct}} =  \frac{\delta c_{\phi ,1}}{2}\int d\tau (\partial_i \phi)^2 + \frac{\delta c_{\omega^2\phi ,0}}{2}\int d\tau \dot{\phi}^2\,,
\end{equation}
with 
\begin{align}
    \delta c_{\omega^2\phi ,0} =&\pi^2 r_{s,5}^4 \left(\gamma_E +\log (\pi )+\frac{1}{\epsilon _5}\right)\,, \\
    \delta c_{\phi ,1} =& \frac{\pi^2 r_{s,5}^4}{16}\left( \gamma_E +\log \left(\pi \right)+\frac{1}{\epsilon _5}\right) \,.
\end{align}
The renormalized action in the  $\overline{\rm MS}$  scheme is then given by  
\begin{equation}
    S^{\text{ren}} =  S_{\text{tidal}} +S^{\text{ct}} =  \frac{\bar{c}_{\phi ,1}}{2}\int d\tau (\partial_i \phi)^2 + \frac{\bar{c}_{\omega^2\phi ,0}}{2}\int d\tau \dot{\phi}^2\,, 
\end{equation}
where the renormalized Love numbers  are
\begin{equation}\label{eq:renlovenumbers}
    \begin{split}
        \bar{c}_{\omega^2 \phi,0} &= {-}\pi^2 r_{s,5}^4   \left(2 \log \left(\mu r_{s,5}\right)+\frac{11}{6} \right)  \,,\\
        \bar{c}_{\phi,1} &=-\frac{\pi^2 r_{s,5}^4}{16}\left(2 \log \left(\mu  r_{s,5}\right)+\frac{5}{6}-\log {16}\right)\,,
    \end{split}    
\end{equation}
producing  therefore a finite $S^{\text{ren}}$. 

\end{widetext}

\bibliography{short.bib}

\end{document}